# Molecular-scale Integration of Multi-modal Sensing and Neuromorphic Computing with Organic Electrochemical Transistors

Shijie Wang[1]†, Xi Chen[2]†, Chao Zhao[1], Yuxin Kong[3], Baojun Lin[1], Yongyi Wu[1], Zhaozhao Bi[1], Ziyi Xuan[1], Tao Li[1], Yuxiang Li[3], Wei Zhang[4], En Ma[4], Zhongrui Wang[2]*, Wei Ma[1]*

Shijie Wang[1], Chao Zhao[1], Baojun Lin[1], Yongyi Wu[1], Zhaozhao Bi[1], Ziyi Xuan[1], Tao Li[1], Wei Ma[1]*

[1]State Key Laboratory for Mechanical Behavior of Materials, Xi'an Jiaotong University, Xi'an, China.

Xi Chen[2], Zhongrui Wang[2]*

[2]Department of Electrical and Electronic Engineering, The University of Hong Kong, Pokfulam, Hong Kong.

Yuxin Kong[3], Yuxiang Li[3]

[3]School of Materials Science and Engineering, Xi'an University of Science and Technology, Xi'an, China.

Wei Zhang[4], En Ma[4]
[4]Center for Alloy Innovation and Design (CAID), State Key Laboratory for Mechanical Behavior of Materials, Xi'an Jiaotong University, Xi'an, China.

†These authors contributed equally to this work.

*Corresponding author. Email: zrwang@eee.hku.hk; msewma@xjtu.edu.cn

**Abstract:** Bionic learning with fused sensing, memory and processing functions outperforms artificial neural networks running on silicon chips in terms of efficiency and footprint. However, digital hardware implementation of bionic learning suffers from device heterogeneity in sensors and processing cores, which incurs large hardware, energy and time overheads. Here, we present a universal solution to simultaneously perform multi-modal sensing, memory and processing using organic electrochemical transistors with designed architecture and tailored channel morphology, selective ion injection into the crystalline/amorphous regions. The resultant device work as either a volatile receptor that shows multi-modal sensing, or a non-volatile synapse that features record-high 10-bit analog states, low switching stochasticity and good retention without the integration of any extra devices. Homogeneous integration of such devices enables bionic learning functions such as conditioned reflex and real-time cardiac disease diagnose via reservoir computing, illustrating the promise for future smart edge health informatics.

## Introduction

The unprecedented advancement progress in artificial intelligence (AI) is revolutionizing a broad spectrum of applications such as computer vision, natural language processing, and industrial automation in recent year. The traditional AI hardware employs physically separated information sensing, processing, and memory architecture, which suffers from large energy and time overhead due to frequent data shuttling between physically separated hardware modules and sequential analog-digital conversion[1-3]. Biological nervous systems (BNS), however, significantly outperforms artificial neural networks (NNs) running on conventional silicon hardware in terms of its high energy-area efficiency. For example, the power consumed in the human brain for the remarkably error-tolerant processing of spatiotemporal information based on spike coding is only around 15 W[4,5]. Most significantly, the collocation of sensing-processing-memory functionalities equip BNS with unparalleled intelligence and efficiency, as manifested by complicated bio-behaviors like conditioned reflex. Therefore, bionic learning has a long way to go for efficient hardware implementation, especially for edge AI where real-time sensing, processing and memory are subjected to tight power and form factor constraints.

The diverse functions of building blocks of BNS, especially the receptors in the peripheral nervous system (PNS) for sensing as well as synapses and neurons in central nervous system (CNS) for signal processing, makes hardware implementation of bionic learning challenging. To satisfy the performance demands of BNS, heterogenous module integration has been employed for artificial olfactory[6], tactus[7] and gesture recognition[8]. However, the sensors and processing cores in these systems are still physically separated and structurally different, impacting fabrication compatibility, integration density and conductance matching as the device dimensions are scaling down. Recent development of in-sensor computing on a single device using two-dimensional materials[9,10] is one step closer to bionic learning; however, such devices lack non-volatile memory for signal processing. Phase-change materials (PCMs)[3,11] and redox memristors (RRAMs)[1,12] have been employed for in-memory computing, but show no sensing capabilities. Until now, high-performance devices for homogeneous hardware implementation of artificial BNS is still unavailable[13].

Recently developed solution-processed organic electrochemical transistors (OECTs) with organic mixed ion-electronic conductors (OMIECs) components, ion-mediated mechanism, wet operating environment and low power consumption demonstrate either sensing or analog memory[14] capabilities similar to the biological neurons. However, the majority of OECTs reported so far target the sensing of a certain type of signal such as chemical[15,16] and electrophysical[17,18] stimuli, an OECT based multi-modal sensing unit that emulates the biological polymodal receptor, such as TRPV1 of human beings[19] is in urgent need for complicated machine learning process. Moreover, to achieve fast response, the injected ions shall freely diffuse back to the electrolyte and reinstate the channel to its initial state in aforementioned OECTs, making it impossible for non-volatile conductance modulation. On the contrary, by confining the ionic drift through enforced gate-channel open circuit, non-volatile OECTs has been demonstrated with more than 500 analog states and 100 second state retention[20-22]. However, such open-circuit condition can only be met when heterogeneous devices (e.g. conductive bridge memories) were integrated with OECTs (see Table S1). In contrast, tunable volatile/non-volatile behaviors of OECTs can be translated to homogeneous integrated hardware with fused multi-modal sensing-memory-processing functions as well as associated advantages (e.g. compactness, fabrication and conductance compatibility, and bio-combability)[23], as illustrated in Fig. 1a, which are yet to be demonstrated mainly due to the contradictory ion kinetic in volatile and non-volatile OECTs. In this work, we propose a vertical traverse architecture together with designed crystalline-amorphous channel that can be selectively doped by ions to enable dual-modes operation of OECTs, which leads to homogeneous and efficient BNS hardware with colocation of sensing, processing and memory. We then exploit our multi-functional design in bionic learning examples, including environment-aware conditioned reflex and near-sensor computing, to illustrate the potential of our new scheme in addressing the demanding challenges facing by bionic learning hardware.

**Results**

*Design strategies for dual-mode OECT*. To satisfy the completely opposite requirement for retention of channel conductance in volatile/non-volatile OECTs, different device architectures along with the extra devices interfacing the transistor gate and channel were applied. Typically, non-volatile OECTs are achieved by compensating the counterions in electrolyte, or by increasing the hopping energy barrier of ions through bonding effects or channel microstructure control[24]. However, for these devices, the electric field between the channel and the gate reverses when the gate voltage is removed, which still yields a driving force of the ions to drift out of the channel when extra devices are absent[25,26]. To prevent this and realize "*true non-volatility*", the energy barrier has to be large enough, ideally, the ions shall firmly trap within the compact, ordered and bulky side chains rather than simply be blocked by the surrounding crystallites. Moreover, the reversed electric field has to be weak enough, which indicates the importance of the long channel depth. For volatile OECTs, on the other hand, ions prefer to inject/diffuse into/out of the amorphous region. Therefore, increasing the dimension of crystalline domains will inevitably reduce the volume capacitance (C*) and ion mobility, making it difficult to capture the faint signals such as electrophysical and thermal stimuli. To achieve dual operation modes in a single OECT, we propose a new scheme with the following features: 1) Device architecture with large channel depth to flatten the electric field within the channel, meanwhile the large depth/length (*d/L*) geometric ratio to compensate the loss of C* and ion mobility for highly sensitive multi-modal perception; 2) the crystallization control to ensure the ions can be trapped within the ordered and compact molecular chains, and shuttles among the amorphous chains easily, while avoiding the generation/existence of tiny cations that can permeate the crystal and neutralize the channel.

*Vertical traverse architecture design.* We propose a vertical traverse OECT (v-OECT, Fig. 1b) architecture gated by [$EMIM^+$][$TFSI^-$]:PVDF-HFP ion gel or aqueous solution, with a naturally formed crossbar structure. In our v-OECT, the channel "length" $L$ is 40 nm to 60 nm, as determined by the film thickness, whereas the channel "thickness" $d$ is 100 μm, therefore yielding a super high *d/L* ratio of around 2,000, which guarantees an ultra-high sensitivity artificial receptor. More importantly, the large *d/L* ratio leads to much smaller electric potential gradients along the *d*-direction, which effectively prevents the trapped ions from drifting out of the channel after removing the gate voltages, as corroborated by the non-volatile performance of p-OECTs with different channel thickness (Fig. S7). As shown in Fig. 1g, for p-OECT with ultra-thick channel (~2 μm), the relative quantity of $F^-$ (the characteristic element of [$TFSI^-$]) in bulk channel can be much higher than that at the electrolyte-channel interface. This suggests that only a tiny number of anions travel back to the electrolyte after the withdraw of gate voltage, leaving a narrow neutral interface, whereas the majority of trapped ions are still confined inside the bulky crystals (domain size ~ 30 nm, evaluated from cryo-EM image and GIWAXS data, Fig. 1c, S4-S6) and thus leads to the non-volatile behavior. Furthermore, the small $L$ promises fast response and wide conductance dynamic range, a key demand for both sensing and memory programming.

*Crystallization and electrode process control.* We studied the effect of channel crystallinity in volatile and non-volatile behavior of v-OECT by tracing the fingerprint of ions in virtue of *in-operando* UV-Vis absorption spectra and synchrotron radiation X-ray scattering[27,28]. The measurement setup is shown in Fig. 1e inset where the OMIEC (PTBT-p) films were annealed at different temperatures to tune the crystallinity (Fig. 1c, S4-S5). As shown in Fig. 1e and S2, the normalized 0-1 absorption peaks, fitted into the intensity-potential plot for annealed conditions, appear with significantly different slopes, which corresponds to the anion doping in amorphous (stage I) and crystalline (stage II) regions, respectively. The breakpoint between stages appears earlier with stronger absorbance along with the enhancement of crystallinity, this phenomenon agrees well with the increasing crystal ratio and the associated wider memory window (Fig. S4-S5). For both as-cast and 200 °C annealed films, [$TFSI^-$] can sufficiently *p*-dope the OMIEC under high gate potential (HGP, -1.5 V), as indicated by the disappeared 0-1 and emerged polaron signals. However, when the "gate" is grounded, the absorption signals of the as-cast film relax back to the initial state within 5 s, while the 200 °C annealed film maintains in *p*-doped state, which directly proves its *"true non-volatility"*. When the same samples are pre-biased at low gate potential (LGP, -0.7 V), the OMIEC can be partially *p*-doped and rapidly recover to its neutral state when "gate" is grounded. Together with the first stage information in Fig. 1e, this indicates that the LGP can dope the amorphous region, but in a volatile manner. Moreover, X-ray scattering directly traced the anions in a crystalline channel (Fig. 1h), the lamellar stacking obviously expands from 1.39 nm to 1.53 nm only when sufficient HGP is supplied, suggesting that anions trapped among the ordered and compact glycol side chains and can de-trap only when sufficient positive potential is applied. This trapping/de-trapping process corresponds to the reversible potentiation/depression (P/D) of v-OECT synapses. We quantify the energy barrier to embed [$TFSI^-$] into the crystalline glycol side chains is around 0.8 eV when polarizable electrode (Au) was used, as revealed by the breakpoint potential of 200 °C annealed sample (Fig. 1e). When non-polarizable Ag/AgCl was used as the gate electrode, only volatile behavior appears mainly because the counter-ions ([$EMIM^+$]) are more inclined to permeate and neutralize the channel rather than form the stable electric double layer and fixed near the polarizable gate electrode (Supplementary text, Fig. S8-S9). Therefore, as summarized in Fig. 1i, with careful control of the crystallinity and electrode process, the volatile/non-volatile properties can be simultaneously achieved on a single cv-OECT.

**Volatile receptor behaviors.** We applied LGP to evaluate the volatile performance of v-OECTs for multi-modal bio-sensing function (Fig. 2a). The signal amplification capability of OECT strongly depends on the channel dimension and the figure of merit ($\mu C^*$). Fig. 2b shows the volatile transfer curves at different channel crystallinity. The $\mu C^*$, and thus the transconductance of the v-OECTs, increases with the crystallinity (Fig. 2b-c and S10-S11). Compared to p-OECT, hole mobility ($\mu$) of v-OECT benefits from the crystallinity when gated by both ion gel and aqueous solution (Fig. 2c and S19, supplementary text), indicating that ultra-short channel restrained the heterogenous swelling effect in crystalline OMIECs (Fig. S19-S20, supplementary text)[29]. The crystalline v-OECT annealed at 200 °C (called cv-OECT) shows a normalized peak transconductance of $g_m/V_{DS}$ = 27 mS/V, an on/off ratio of $5 \times 10^5$, and a subthreshold swing of SS = 65 mV/dec, The SS is close to the thermodynamic limit (59.6 mV/dec) in a broad subthreshold regime around $V_{GS}$ = 0 V, much superior than those of the p-OECTs (121 mv/dec) and reported vertical electrolyte gated transistors such as PDPP3T based v-EGOFET (90.5 mv/dec)[30]. As a key parameter, the on/off ratio of cv-OECT can be up to $8\times10^6$ by tailoring the channel geometry (Fig. S13), a record-high value in OMIEC-based transistors. The ion permeable OMIEC-based channel also enables devices to work in aqueous environment (0.1 M NaCl, Fig. S14-S15) without performance degradation. As for transient behavior, the on/off time of cv-OECT calculated by single exponential fitting is 6.67 ms and 3.20 ms, respectively (>150 Hz, Fig. 2d), much shorter than that of p-OECT (> 1s, Fig. S11) because of the smaller doping area of v-OECT, which also mitigates the instability issue. Hence, the cv-OECT maintains a high on/off ratio of $10^6$ and identical switching speed after 30 minutes cycling in air (Fig. 2d).

Fig. 2e and Table S2 summarize the performance of several state-of-the-art electrolyte gated transistors (EGTs) in terms of their SS at 0 V and on/off ratios, which are benchmarks of power efficiency and amplification capability. The ultra-low SS with high on/off ratio, together with the fast response, make cv-OECT a great choice to emulate the power-efficient multi-sensory biological receptor, an integral part of bionic learning. As a demonstration, the local ion concentration changes of *Mimosa pudica* and *Venus flytrap* caused by light and mechanical stimuli were clearly revealed by the cv-OECT (Fig. 2f and Fig. S21), without the need of complex amplifying circuits. Moreover, as a classical benchmark, ECG recording was carried out by cv-OECT working in subthreshold regime with a much lower energy consumption (< 1 μW) compared with PEDOT:PSS-based OECTs (usually > 500 μW) (Fig. 2g and Fig. S22)[17]. More importantly, we also demonstrate that cv-OECT provides a unified solution for multi-modal sensory NNs in edge computing[31] (Fig. 2h and S23). As a proof of concept, cv-OECT shows gustation and temperature sensation with the high normalized response (NR) of 19 %/dec and 3.2 %/°C, respectively. The ultra-short channel together with the trap filling effect also promises its application to artificial vision (Fig. S23e)[32].

**Non-volatile synaptic behaviors.** The device architecture of non-volatile synaptic v-OECT

is exactly the same as that of volatile ones (Fig. 3a), except a higher gate operation voltage. As shown in Fig. 3b, when gated by the ion gel, the cv-OECT conductance change can be long-term modulated by applying gate pulses with amplitudes larger than |-0.8| V. Below this, cv-OECT is volatile (Fig. S24). The "*truly non-volatility*" is also reflected by the transfer curves that show ideally centrosymmetric hysteresis (Fig. 3c), providing a memory window of 2.1 V[33]. We evaluated the state retention of cv-OECT (Fig. 3d) with the gate grounded (a prerequisite for large-scale homogeneous integration in NNs), the conductance of cv-OECT was switch between eight analog states and state is capable to maintain for more than 20,000 s in ambient air with a conductance drift coefficient of $\gamma \sim 0.008$ (Fig. 3d inset and Fig. S25), comparable to the state-of-the-art heterostructure-based PCMs[11].

Energy and time efficient analog in-memory computing requires linear, symmetric and precise conductance update as well as a large number of states in a wide dynamic range[34]. As depicted in Fig. 3f, cv-OECT shows 1,024 (10-bit) distinct states over a wide dynamic conductance range in LTP. More importantly, the number of bits can be further increased by reducing the programing pulse width/amplitude (Fig. 3b and S24), making cv-OECT effectively a "*truly analog*" device. This advantage benefits from the low intrinsic conductance of our OMIEC and the large transconductance of cv-OECT. As a result, the device is immune to memory/computing errors due to write/read noise and conductance drift even under a large density of conductance states. While the write noise of cv-OECT can be well confined under both voltage and current control, as a non-linearity of $\upsilon_P/\upsilon_D = 0.20/1.63$ and a high signal-to noise level of $(\triangle G_{DS}/\sigma)^2 \approx 179$ were achieved for standard 50-states programming (Fig. 3f and S26-27), where σ is the standard deviation of the conductance update. When current pulses (±200 nA, 20 ms) were applied, cv-OECT shows a $(\triangle G_{DS}/\sigma)^2 \approx 110$ together with a low cycle-to-cycle variation ≈ 0.49% during 2,000 P/D events in 50 cycles (Fig. 3e and S27), and the one-to-one correspondence was established between each state in P/D process. Note the width of current pulses to program cv-OECT is far shorter than that of previous report (mostly >1 s)[35], making current-driven cv-OECT capable for real-time online training of NNs. In addition, cv-OECT also provides a general platform for mode-switchable transistors (e.g. PDPP3T based device, Fig. S29). The long retention, low switching stochasticity and large number of analog conductance states make cv-OECTs promising for analog in-memory computing to process sensory information feedback from receptors in real time.

Beyond LTP, we demonstrated spike-timing dependent plasticity (STDP), a more fundamental local learning rules practiced by the human brain[36]. The STDP synapse contains two identical cv-OECTs, forming a 1-transistor-1-memristor (1T1R) architecture (Fig. 3g inset, Fig. S30). The source of "***T***" is in series with the gate of "***R***" and the $I_{DS}$ of "***R***" is monitored by a read pulse ($V_{read}$). Paired nervous impulses (named pre-spike/post-spike) are applied to the synapse with a time delay (Δt). The post-spike (HGP) drops on the gate of "***R***" decreases with the increase of the Δt, the same as the non-volatile conductance change of "***R***" because "***T***" is volatilely tuned by the pre-spike (LGP). This relation is well fitted to a single exponential function (Fig. 3h and S31) with a time constant of ~60 ms, similar with its biological counterpart[36]. Some unique advantages of our hardware implementation of STDP includes: 1) Non-volatile, analog, and highly accurate conductance tunning compared to other emerging electronic synapses[37]. 2) The large off state resistance of "***T***" channel is less prone to conductance drift caused by sneak gate current[21]. 3) No need of any heterogeneous integration (e.g. diffusive and drift memristors) or complex pulse engineering[38], which provides the building block of homogeneous bio-plausible spike-neural networks (SNN)[2]. Here, we monolithically integrated cv-OECTs on the same chip to implement SNNs equipped with the hardware-encoded supervised STDP learning rule (Fig. 3g). The array consists of 18 cv-OECTs, (9 "***T***"s and 9 "***R***"s, Fig. 4a and S32). Every two cv-OECTs on the same row form a STDP synapse. The input to the SNN is mapped to the pulse width of spikes. Right after that, pre-spikes (-0.8 V, 100 ms) will be applied to the gate of "***T***" of different rows and the output currents will be integrated by post neurons. The neuronal states are then compared with a teacher signal to decide whether a set or reset (-2 V or +2 V, 2 ms) post-spike will be applied to the drains of the "***T***"s in different columns to potentiate/depress the synapses via the aforementioned STDP rule. As a result, neurons that fire together tend to wire together, a manifestation of the Hebbian learning rule. When the post-spike/pre-spike are completely overlapped, the identical network can also work as an ANN with parallel programming capability thanks to the 1T1R architecture. We compare the simulated performance of such a homogeneous single-layer SNN or ANN with the alternative implementations using RRAMs in classifying MNIST handwritten digits (Fig. 3h) based on the experimentally measured non-linearity, cycle-to-cycle and device-to-device variation of cv-OECT arrays (Fig. S32). The resultant SNN shows a classification accuracy of ~89% (Fig. 3i), comparable to the result of the ANN based on the homogeneous cv-OECT array (~91%), which is higher than that of 6-bit heterogenous RRAM based SNN (~83%) and ANN (~87%), illustrating its potential for energy-efficient and error-tolerant bio-inspired computing.

**Integrated functions for fused bionic sensing-processing.** Leveraging the dual operation mode of cv-OECT, we mimicked the fused sensing-processing function of BNS. Firstly, ***Conditioned reflex***, a commonly seen behavior that helps bio-creatures to better survive the evolution, was demonstrated using a circuit similar to that of STDP (Fig. 4a-b and S30), where the "***T***" is gated by aqueous solution with varied ion concentration and species to perceive information while "***R***" is gated by ion gel for memory. A DC voltage ($V_{dd}$ = -0.8 V) is supplied to the drain of "***T***". When the "*Bell"* (+0.4 V) signal spikes, $V_{dd}$ has a negligible influence on the conductance (memory level) of "***R***" due to the large off state resistance of the cv-OECT. Once the "*Food*" signal (-0.8 V) spikes together with the *"Bell"* signal, the $G_{DS}$ of "***T***" can be temporally switched by LGP, making the $V_{GD}$ higher than -0.8 V so that the conductance of "***R***" can be tuned in a non-volatile manner. More importantly, the learning rate strongly depends on learning environment, the effective gate potential of "***T***" is sensitive to the ion concentration and species due to Nernst potential, making memory level of "***R***" correlated to the logarithmic of anion activity (Fig. 4c inset). As such, the sensory information was encoded into the memory in a bio-plausible manner which can be further extended to complex fused chemical sensing-processing at the edge.

Finally, as a proof of concept, real-time *Cardiac disease diagnose* was achieved via reservoir computing (RC) by the all-homogenous integration "sensing-processing" with identical cv-OECTs. As shown in Fig. 4d, 12-lead ECG signals of five kinds of cardiac diseases were captured by a 12×1 cv-OECT based receptor array, which also acted as neurons or computing nodes of a dynamic reservoir. The captured ECG signals were leaky integrated on receptor the $I_{DS}$ output of which were then sampled and used as feedback to a 156×5 cv-OECT based ANN readout map for classification. The simulated diagnoses accuracy based on the experimentally calibrated device models reached 100% after 700 training epochs (Fig. 4e, see method). Note that the aforementioned multi-modal sensing, such as body temperature, fluid monitoring and virus detection, can be seamlessly integrated into the reservoir, leading to portable and efficient edge-AI hardware for healthcare applications.

In summary, the innovations in device architecture and channel microstructure equipped a single cv-OECT with multi-modal sensing and non-volatile analog memory simultaneously. We demonstrated bionic SNNs chips leveraging hardware-encoded STDP learning rules, and the fused sensing-processing such as environment-aware conditioned reflex based on homogeneous integration of cv-OECTs, which may help to solve the performance and integration issues in future bionic edge-AI hardware.

**Acknowledgments:** X-ray data was acquired at beamlines 7.3.3 at the Advanced Light Source, which is supported by the Director, Office of Science, Office of Basic Energy Sciences, of the U.S. Department of Energy under Contract No. DE-AC02-05CH11231. The authors thank Chenhui Zhu at beamline 7.3.3 for assistance with data acquisition and acknowledge the financial support by NSFC (21704082, 21875182), Key Scientific and Technological Innovation Team Project of Shaanxi Province (2020TD-002) and 111 project 2.0 (BP2018008).

**Author contributions:**
S.W. and W.M. designed the experiments. Y.L and Y.K synthesized the OMIEC. S.W. collected and analyzed data. Z.W. and X.C. designed the network simulations. All authors discussed the results and contributed to the manuscript preparation. S.W., C.Z., W.M., Z.W., W.Z. and E.M. wrote the manuscript.

**Competing interests:** Authors declare that they have no competing interests.

**Data and materials availability:** The data that support the findings of this study are available from the corresponding authors upon reasonable request.

**Supplementary Information**

Materials and Methods

Supplementary Text

Figs. S1 to S33

Tables S1 and S2

References (39–62)

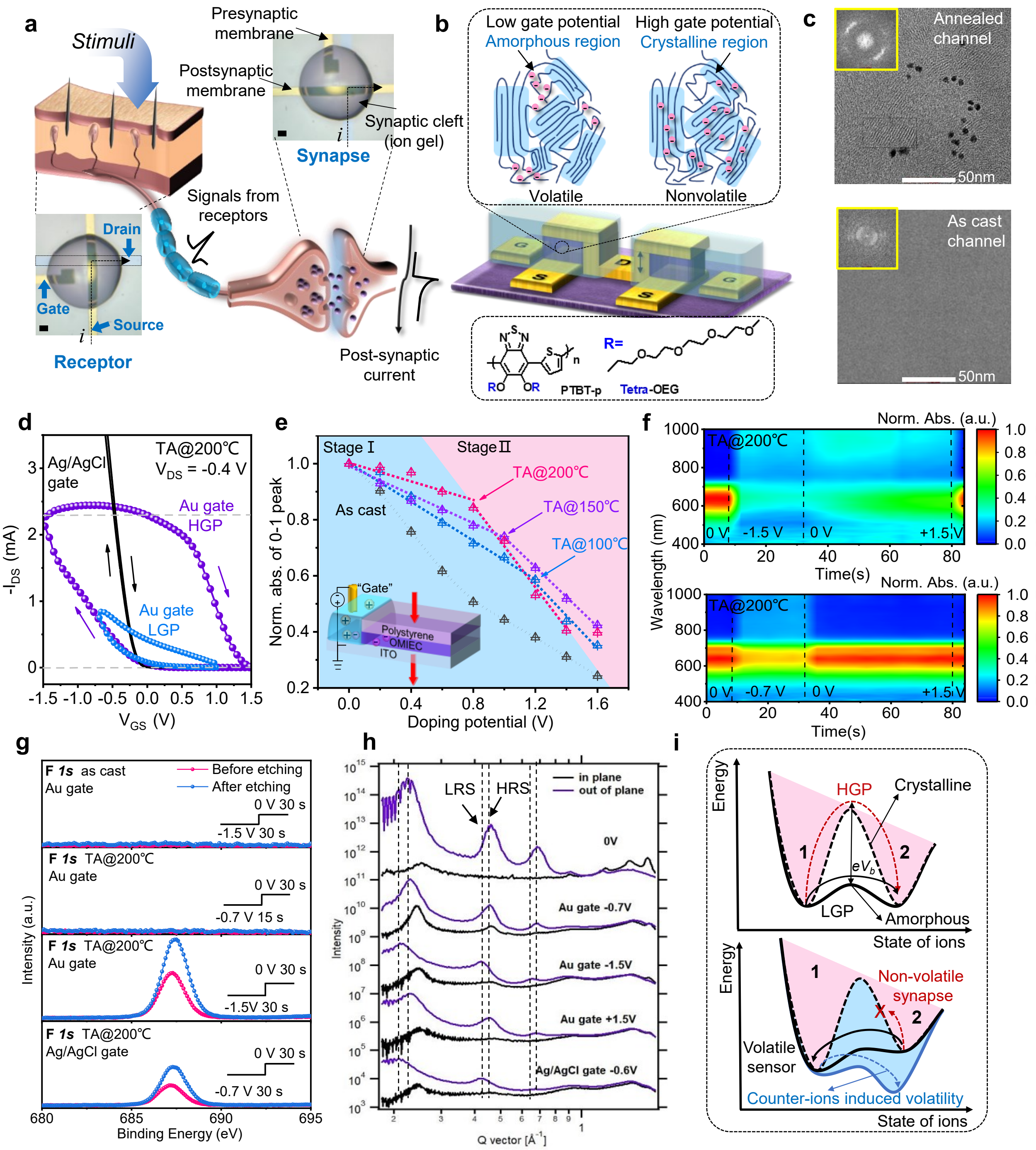

**Fig. 1 | Design of the mode switchable cv-OECT.** **a**, Comparison between biological nervous system and cv-OECT based artificial nervous system, where cv-OECT can act as both volatile receptor and non-volatile synapse. Optical micrographs display the top view of a v-OECT (scale bar: 100μm). **b**, Device architecture of v-OECT, two dashed boxes show the ion contribution in volatile/non-volatile mode and chemical structure of PTBT-p, respectively. **c**, Cryo-EM images of the 200 ℃ annealed and as-cast PTBT-p films **d**, Transfer curves of cv-OECT with polarizable/non-polarizable gate electrode. **e**, Normalized 0-1 absorbance as a function of doping potential; inset shows the setup for UV-vis measurement. Stages I and II corresponds to the doping of amorphous and crystalline regions, respectively. **f**, Time resolved UV-vis spectra of channels correspond well with the device performance. **g**, XPS spectra of as-cast and annealed p-OECT channels doped at LGP and HGP. Pink and blue

lines are the signals from [TFSI⁻] before and after 30 nm-etching. **h**, GIWAXS profile of annealed channels doped at LGP or HGP. Reversible displacement of (100) peak suggests that the anions firmly embed among the glycol side chains in the crystalline region. **i**, Schematic explaining the mode switching mechanism. Special channel dimension and crystallization provide a high barrier *eVb* between the two ionic states (1 and 2), resulting in non-volatile behavior; LGP can only inject the ions into amorphous regions and lead to volatile behavior; When non-polarizable electrode was used, the counter-ions are more easily to permeate and neutralize the channel in virtue of reversed electric field, making the conductance volatile.

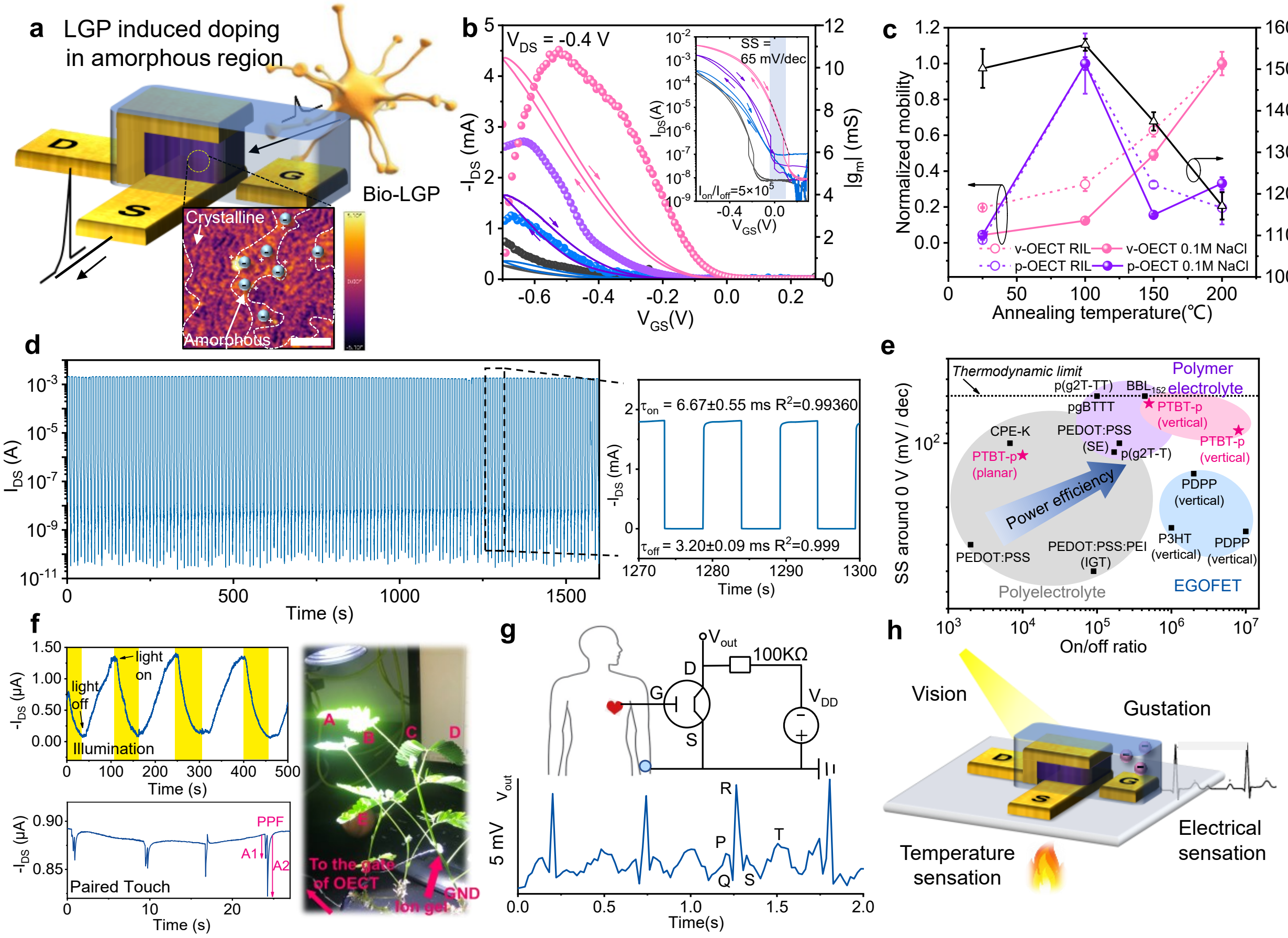

**Fig. 2 | Volatile receptor behavior.** **a**, Schematic illustration of cv-OECT acting as a volatile receptor. Black box shows the KPFM phase image and the anion distribution in cv-OECT channel under LGP (Scale bar: 200 nm) **b**, Transfer curves and related transconductance of ion gel gated v-OECT annealed at various temperatures (pink: 200℃, purple: 150℃, blue: 100℃ and black: as-cast); transfer curves (inset) with the $I_{DS}$ plotted on a logarithmic axis show decent SS and on-off ratio (L = 40 nm, W = 100 μm, similarly hereinafter). **c**, Volume capacitance and normalized operando mobility of v-OECT and p-OECT gated by ion gel and 0.1 M NaCl aqueous solution (represent data from n = 3 individual devices). **d**, Stability of cv-OECT tested in ambient environment, high on/off ratio and switching speed can be kept during the cycles. Close-up shows high-speed switching of cv-OECT operated in volatile mode. **e**, Comparison of SS aroun 0 V and on/off ratios for different EGTs. **f**, Recorded $I_{DS}$ as a function of time upon stimulating the *mimosa pudica* by LED light and paired touch, the paired pulse faciltation (PPF) behavior was captured. **g**, Wiring diagram of ECG recording using a cv-OECT based circuit at $V_{DD}$ = -0.8 V and $V_{GS}$ = 0 V. The gate and grounded source electrode were connected to the left chest and right wrist, respectively. Sample trace of acquired ECG signals is shown. **h**, Schematic illustration of cv-OECT acting as a multi-sensory receptor.

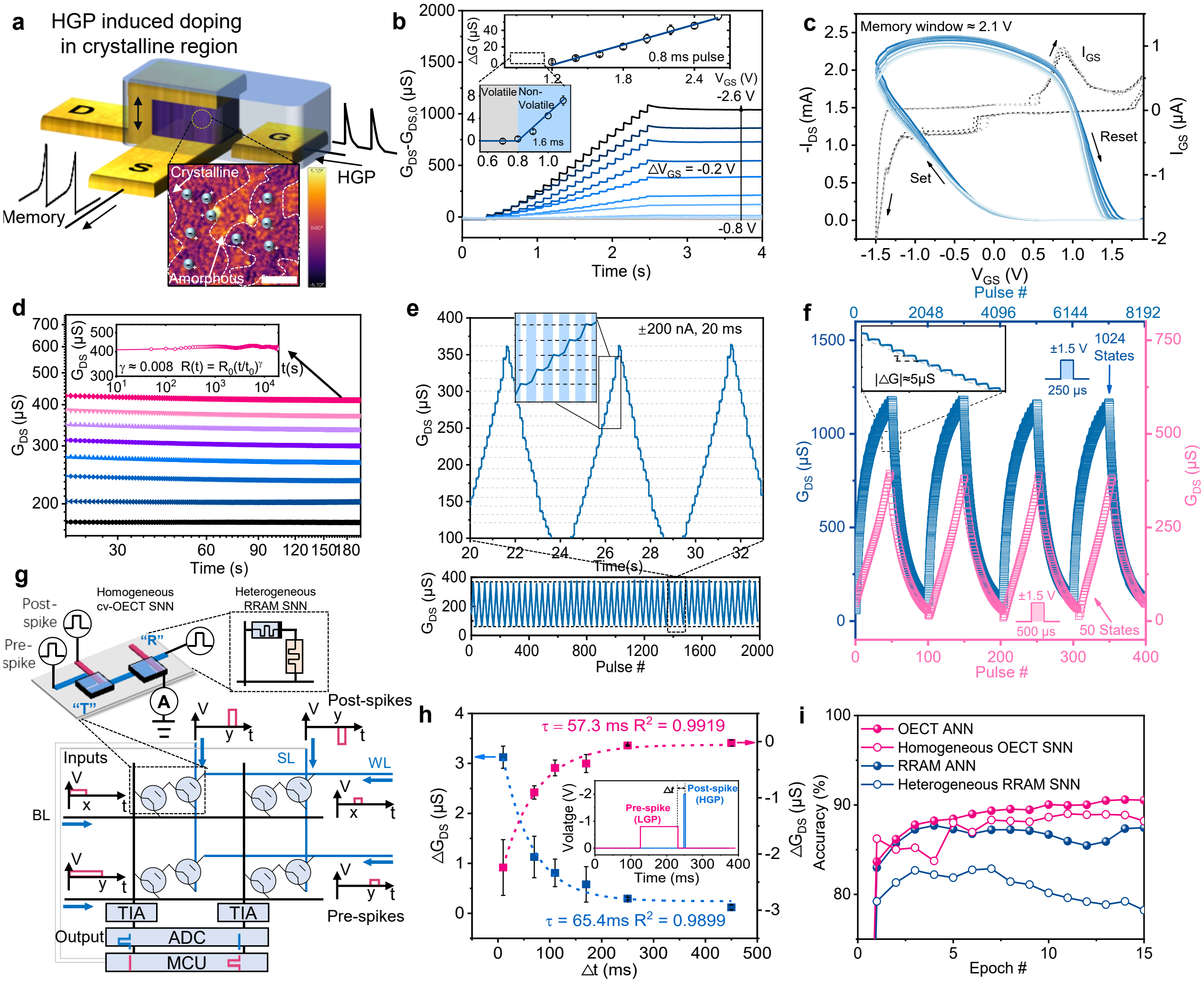

**Fig. 3 | Truly non-volatile synaptic behavior and neural networks.** **a**, Schematic illustration of cv-OECT acting as a non-volatile synapse. Black box shows the anion distribution under HGP (Scale bar: 200 nm). **b**, Non-volatile conduction changes of cv-OECT as a function of gate pulse amplitude. The pulse greater than |-0.8| V can switch the cv-OECT to the non-volatile mode. Signals are filtered to eliminate the read noise and volatile spikes. **c**, Transfer curves of cv-OECT with polarizable gate under HGP. The centrosymmetric drain and gate current with obvious hysteresis demonstrate the nonvolatile nature. **d**, State retention of eight analog states (with gate grounded, similarly hereafter). Inset shows that each state can be maintained for longer than 20,000 s with low drift. **e**, Cyclic LTP under current control (2,000 pulses, ±200 nA, 20 ms). Upper panel shows three reproducible LTP with linear, symmetrical programming and one-to-one correspondence. **f**, LTP of cv-OECT under voltage control. The blue/pink boxes display 1,024/50 analog states with low stochastic. **g**, Architecture of the homogeneous SNN based on cv-OECT. The inset shows a homogeneous 1T1R unit that contains two identical cv-OECTs, this unit is usually implemented by homogeneous integration of RRAMs. **h**, Plot of the conductance change in STDP and the time delay (Δt) between pre-/post-spike. **i**, Recognition accuracies of the handwritten images from the MNIST dataset using the experimental non-ideal factors of cv-OECTs and RRAMs.

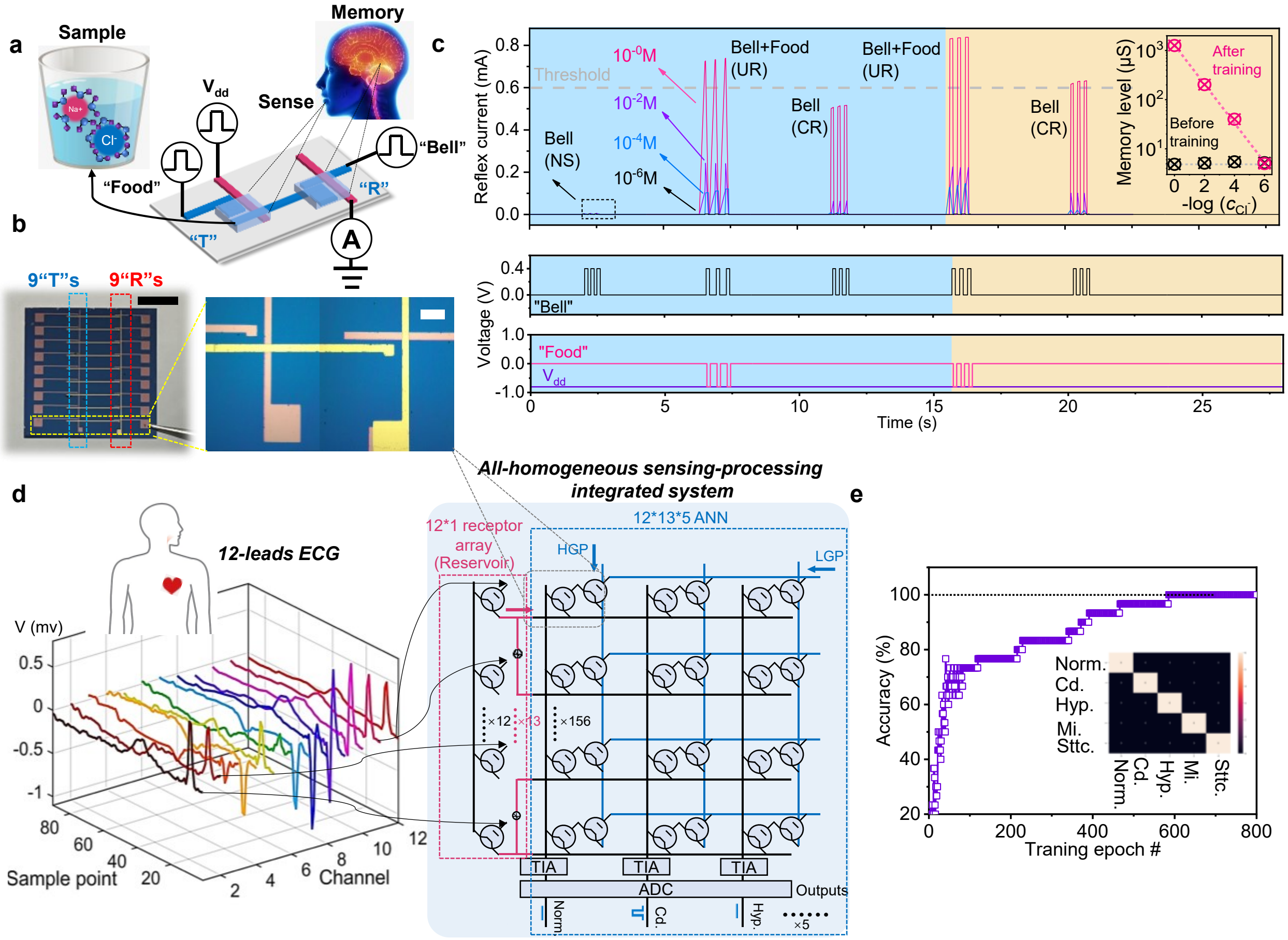

**Fig. 4 | Fused sensing-processing functions.** **a**, Schematic of bionic learning unit that contains two identical cv-OECTs, where "***T***" and "***R***" act as the receptor in PNS and synapses in CNS, respectively. **b**, Image of a 9×2 cv-OECT array which contains 9 "***T***" and 9 "***R***" (Scale bar: 5 mm). Close-up is an image of a bionic learning unit in **a** and Fig. 3g (scale bar: 300 μm). **c**, Learning process and result of the conditioned reflex. **d**, Real-time Cardiac disease diagnose by the all-homogenous integration of cv-OECTs. 12-lead ECG signals of five kind of cardiac patients can be captured by a 12×1 cv-OECT based receptor array, which also acts as a volatile neuron under LGP for reservoir computing. The output of reservoir was delivered to a 12×13×5 cv-OECT based ANN for classification. The grey dashed box shows a 1T1R unit in **b**. **e**, Simulated recognition accuracy of five kinds of ECG waveforms during 800 training epochs. Inset shows the confusion matrix of classification after training.